\documentclass[10pt,twocolumn,letterpaper]{article}

\usepackage[pagenumbers]{cvpr}

\usepackage{xspace}
\usepackage{soul}

\newcommand{\method}{Mimic2DM\xspace}

\usepackage{multirow}
\usepackage{makecell}

\usepackage{amsmath}
\usepackage{amssymb}
\usepackage{amsfonts}
\usepackage{algorithm}
\usepackage{algpseudocode}
\DeclareMathOperator*{\E}{\mathbb{E}}

\definecolor{cvprblue}{rgb}{0.21,0.49,0.74}
\usepackage[pagebackref,breaklinks,colorlinks,allcolors=cvprblue]{hyperref}
\usepackage{cuted}

\title{Learning to Control Physically-simulated 3D Characters via Generating and Mimicking 2D Motions}

\author{
    Jianan Li$^{1}$ \quad
    Xiao Chen$^{1}$ \quad
    Tao Huang$^{2,3}$ \quad
    Tien-Tsin Wong$^{4}$ \\
    \\
    \normalsize $^1$ The Chinese University of Hong Kong \quad
    \normalsize $^2$ Shanghai AI Laboratory \quad \\
    \normalsize $^3$ Shanghai Jiao Tong University \quad
    \normalsize $^4$ Monash University
}

\begin{document}
\maketitle
\begin{strip}
  \centering
  \includegraphics[width=\textwidth]{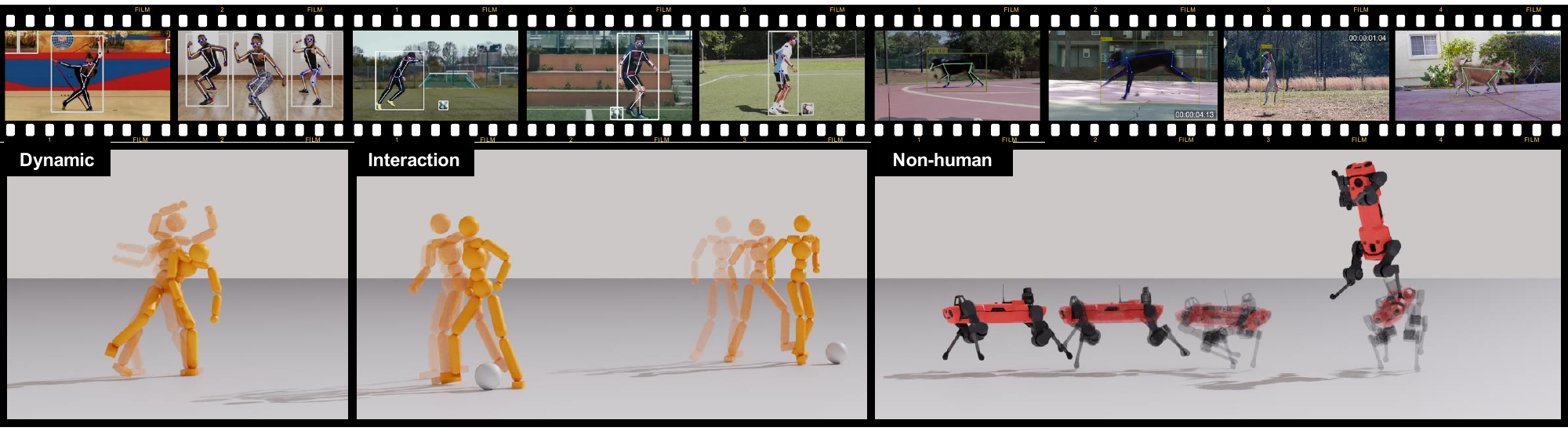}
  \captionof{figure}{The proposed \textbf{\method} effectively learns character controllers for diverse motion types, including dynamic human dancing, complex ball interactions, and agile animal movements, by directly imitating 2D motion sequences extracted from in-the-wild videos.}
  \label{fig:teaser}
\end{strip}
\begin{abstract}
Video data is more cost-effective than motion capture data for learning 3D character motion controllers, yet synthesizing realistic and diverse behaviors directly from videos remains challenging. Previous approaches typically rely on off-the-shelf motion reconstruction techniques to obtain 3D trajectories for physics-based imitation. These reconstruction methods struggle with generalizability, as they either require  3D training data (potentially scarce) or fail to produce physically plausible poses, hindering their application to challenging scenarios like human-object interaction (HOI) or non-human characters. We tackle this challenge by introducing \textbf{\method}, a novel motion imitation framework that learns the control policy directly and solely from widely available 2D keypoint trajectories extracted from videos. By minimizing the reprojection error, we train a general single-view 2D motion tracking policy capable of following arbitrary 2D reference motions in physics simulation, using only 2D motion data. The policy, when trained on diverse 2D motions captured from different or slightly different viewpoints, can further acquire 3D motion tracking capabilities by aggregating multiple views. Moreover, we develop a transformer-based autoregressive 2D motion generator and integrate it into a hierarchical control framework, where the generator produces high-quality 2D reference trajectories to guide the tracking policy. We show that the proposed approach is versatile and can effectively learn to synthesize physically plausible and diverse motions across a range of domains, including dancing, soccer dribbling, and animal movements, without any reliance on explicit 3D motion data. \textbf{Project Website:} \url{https://jiann-li.github.io/mimic2dm/}
\end{abstract}    
\section{Introduction}
Controlling physically simulated characters to perform realistic motion and plausible object interactions remains a fundamental yet challenging problem in computer animation and robotics. Recently, motion imitation techniques have leveraged motion capture (MoCap) data to train physics-based character controllers, achieving impressive results in producing highly dynamic and physically realistic motions on the simulated virtual character~\cite{peng2018deepmimic, lee2021learning, lee2022deep, reda2022learning, hassan2023synthesizing, xu2025intermimic}. However, collecting high-quality 3D MoCap data is costly and labor-intensive, as it requires numerous skilled performers and specialized capture systems.

To address the scarcity of high-quality MoCap 3D data, recent studies have explored exploiting videos as an alternative data source. Most existing methods~\cite{peng2018sfv, yu2021dynamic_movements, zhang2023tennis, mao2024learning} leverage off-the-shelf human motion reconstruction techniques to estimate 3D motions from videos for learning physics-based skills. While advanced training-based estimation methods can achieve remarkable accuracy and realism in reconstructing human motions, their performance heavily depends on extensive high-quality 3D data for training, limiting their applicability in domains with scarce 3D data, such as human–object interactions or non-human motions. Moreover, these methods often result in physically implausible motions due to a lack of physics constraints, which in turn hinders subsequent motion imitation.

In contrast to training on unreliable 3D motions estimated from videos, some studies have demonstrated the possibility of directly utilizing 2D motions extracted from the video footage as supervision, achieving success across various 3D tasks~\cite{chen2019unsupervised, wandt2022elepose, pi2024motion, li2024lifting, kapon2024mas}. This 2D data is highly accessible and can be easily extracted from videos for a wide range of skeletons, including object interactions and non-human (animal) movements. Additionally, 2D keypoint motion detected in videos provides unbiased 2D evidence that accurately reflects the original movements present in the footage. The key challenge when employing 2D data is the missing depth information. While 2D priors combined with geometrical constraints can yield visually plausible 3D poses, the resulting motions are often physically limited and cannot be directly utilized as high-quality data for motion imitation. 

In this paper, we present \method, a generic imitation learning framework capable of acquiring a wide range of complex, physics-based skills, including human-object interaction (HOI) and animal locomotion, by relying solely on widely available 2D motion data extracted from videos. To leverage 2D motion data, we formulate a physics-based 2D motion tracking by unifying 3D reconstruction and physics-based motion imitation into a single reprojection minimization task, which is optimized via reinforcement learning (RL). With the inclusion of physical constraints, the learned policy is able to synthesize physically correct 3D motions by directly imitating from the depth-absent 2D data. Building on this, we introduce a view-agnostic tracking policy. This design not only allows the policy to benefit from diverse viewpoints in the data, thereby learning more realistic 3D motions, but also facilitates easy extension to a multi-view tracking policy for general 3D motion tracking tasks. To enhance training efficiency for single-view motion tracking, we propose an adaptive state initialization strategy along with a reprojection-error-based early termination criterion. Finally, to extend this framework for generative tasks such as novel motion synthesis, we employ a hierarchical control structure that integrates the tracking policy with a 2D motion generator, where 2D motions serve as an interface between the motion generator and the control policy.

We demonstrate that \method can effectively learn a range of challenging physical skills like skillful soccer ball interaction and highly dynamic movements in a robotic dog, using only 2D motion sequences extracted from in-the-wild videos. We further demonstrate that the proposed view-agnostic 2D tracking policy exhibits universal tracking capabilities. As casual videos are likely to be acquired from different or slightly different viewpoints, the single-view policy trained on this data can be effectively extended to a multi-view tracking policy via a view-aggregation mechanism, enabling it to perform 3D tracking. Crucially, we show that, even trained exclusively on 2D data, our approach achieves comparable 3D tracking accuracy to that of 3D motion-trained conventional methods. Furthermore, we highlight the generative potential of our approach by extending it into a hierarchical framework with a 2D motion generator for motion synthesis and conditional control. In this setup, we demonstrate that our proposed autoregressive 2D motion generator surpasses current diffusion-based models in producing the high-quality 2D motion sequences required to effectively guide the tracking policy.

\begin{figure}[t]
    \centering
    \includegraphics[width=0.7\linewidth]{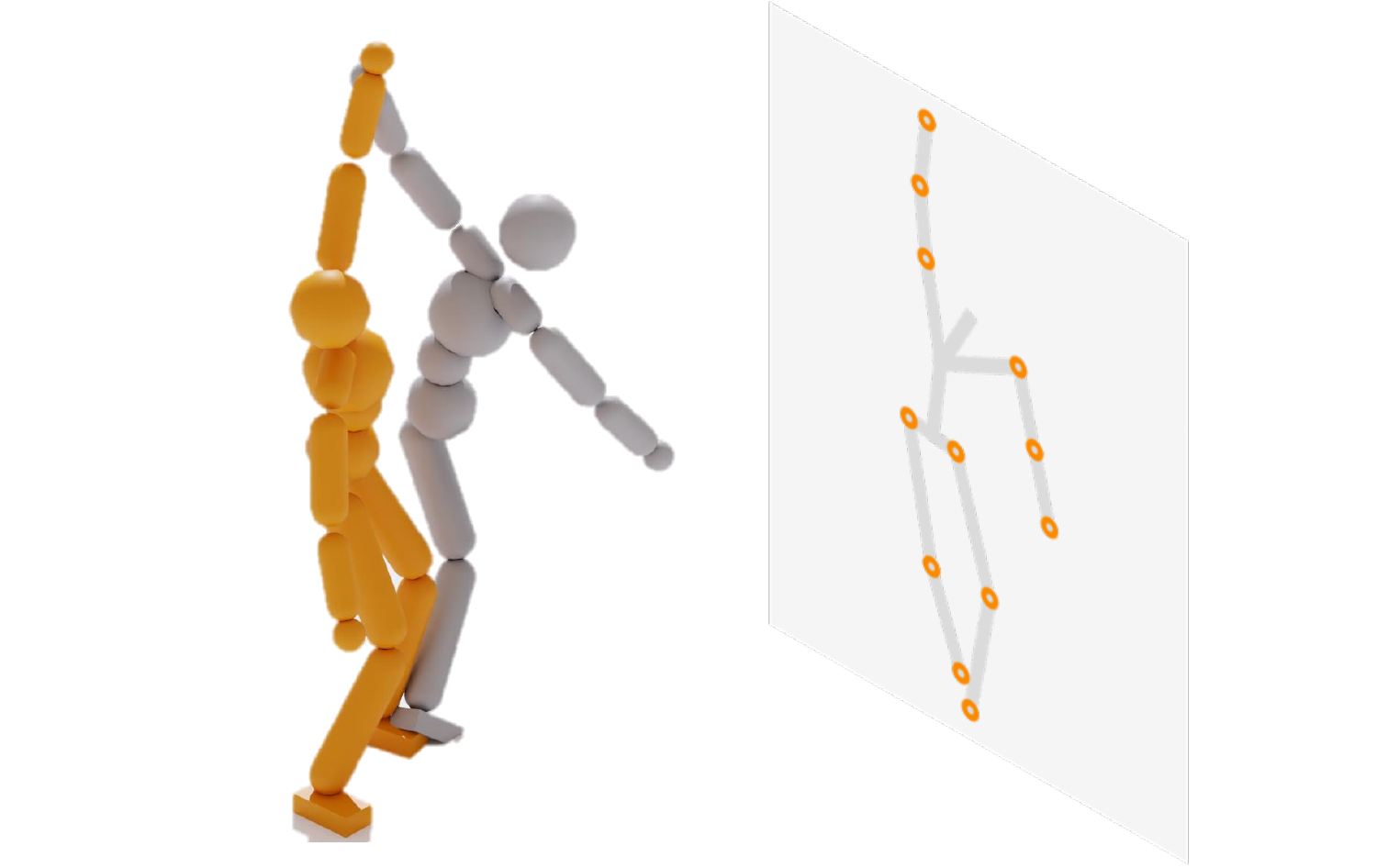}
    \caption{\textbf{Ambiguity caused by 3D-to-2D gap}. When given the 2D tracking controller with only single-view 2D motion as inputs, controllers might struggle to achieve different poses, as illustrated in the figure, the orange colored character and the grey colored character both achieve minimized reprojection error for the given 2D reference.}
    \label{fig:pose_ambiguity}
\end{figure}
\vspace{-2pt}
\section{Related Works}
\subsection{Physics-based Character Control}
Achieving realistic and physically plausible character behaviors is a key goal and challenge in computer animation. To this end, physics simulation is employed to emulate the complex motion dynamics and collision interactions of virtual characters. In early works, physics-based character animation principally focused on locomotion behaviors using traditional optimization-based control strategies combined with some heuristic rules~\cite{raibert1991animation, hodgins1995animating, yin2007simbicon, yin2008continuation, coros2010generalized, de2010feature}. Subsequently, reinforcement learning (RL) was introduced to enable simulated characters to master a wide range of complex skills ranging from basic locomotive skills~\cite{peng2017deeploco, schulman2015high, peng2016terrain, tao2022getup} to proficient sports skills~\cite{chemin2018juggling, liu2018basketball}. However, designing effective reward functions typically necessitates specialized knowledge, and the behaviors generated by reinforcement learning controllers often exhibit irregular motion patterns. To address these issues, MoCap data is employed to train physically simulated character controllers. This can be achieved through explicit motion tracking rewards~\cite{lee2010data, peng2018deepmimic, bergamin2019drecon, park2019learning, luo2023perpetual} or by utilizing implicit motion style rewards derived from a discriminator~\cite{peng2021amp, hassan2023synthesizing, xiaounified}. These approaches enable the learning of more natural and coherent character behaviors.
\begin{figure*}[t]
    \centering
    \includegraphics[width=0.95\textwidth]{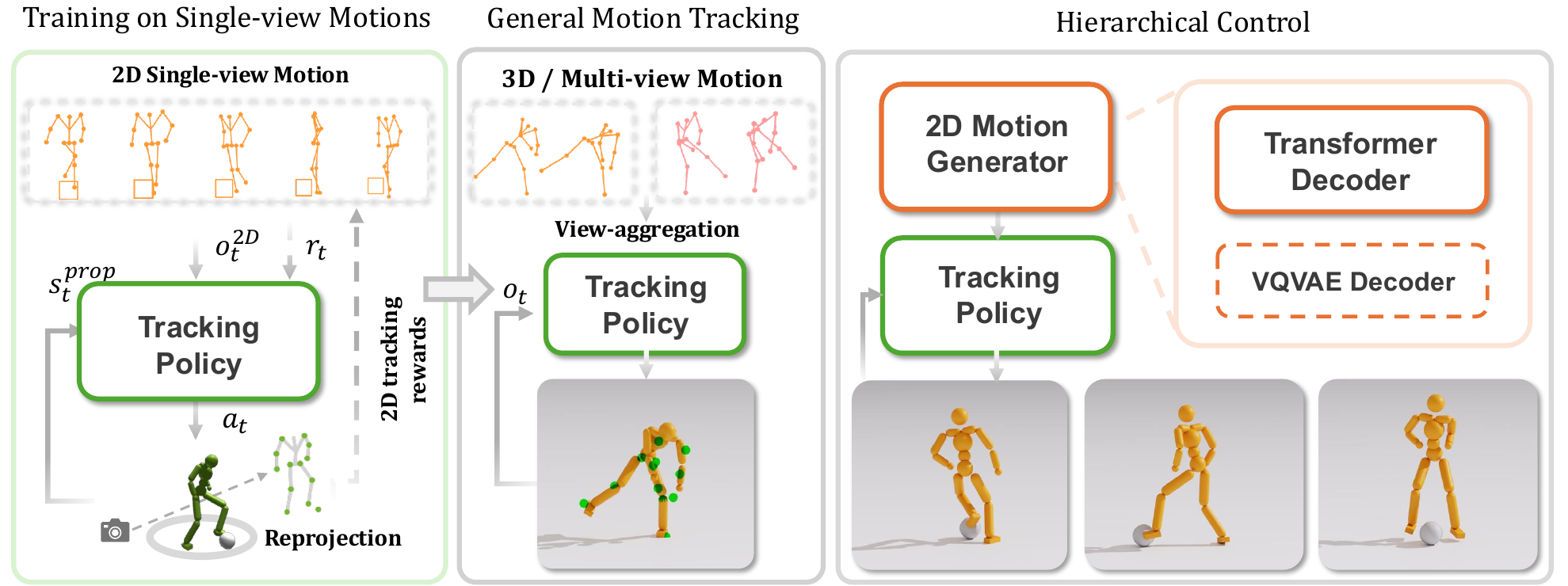}
    \caption{\textbf{Overview of the pipeline}. Our approach \method learns a view-agnostic tracking policy that imitates 2D motion sequences extracted from videos. The resulting policy can be zero-shot adapted to multi-view tracking via view aggregation, and integrated with a 2D motion generator for generative tasks.
    }
    \label{fig:overview}
\end{figure*}

To re-purpose the learned skills for a wide range of downstream tasks, recent work has explored using latent-based generative models, such as VAEs~\cite{won2022physics, yao2022controlvae, luouniversal, yao2024moconvq} and GANs~\cite{peng2022ase, dou2023case, tessler2023calm, juravsky2022padl, juravsky2024superpadl}, to learn reusable motion primitives by mapping motion clips into a low-dimensional latent space. These approaches can efficiently learn a separate high-level policy that controls the pre-trained latent-represented skills for each downstream task. Another line of work focuses on the combination of a universal motion tracking controller with kinematic motion generative models~\cite{tevethuman, li2023object, xu2023interdiff}. This hierarchical control framework also supports versatile, physically plausible motion synthesis and control~\cite{serifi2024robot, tevet2024closd, wu2024human, xu2025intermimic}. However, all of these approaches rely on high-quality 3D MoCap data for training, significantly restricting their applicability and scalability. In contrast, our method is a motion imitation approach that only requires 2D motion data, making it more accessible and versatile.

\subsection{Learning Physically Skills from Videos}
Compared to motion capture (MoCap) data, videos are a more accessible source for learning physics-based skills. Early work by~\citet{vondrak2012video} attempted to reproduce jumping and gymnastic motions in a physics-simulated environment by minimizing silhouette loss from monocular videos. Recent advances in computer vision have enabled the reconstruction of 3D human poses from videos, which can be leveraged to train physics-based character controllers. Initial attempts focused on reproducing individual motion instances in a physics simulator derived from single videos.
For example,~\citet{peng2018sfv} proposed a motion imitation pipeline that tracks 3D poses estimated from video. Building on this,~\citet{yu2021dynamic_movements} advanced the technique by incorporating additional cues such as 2D/3D poses and foot contacts into policy learning, enabling the synthesis of agile motions from long video sequences with dynamic camera movements. To avoid time-consuming physics-based motion imitation for new video clips, \citet{yuan2021simpoe} introduced a real-time physically motion estimation approach SimPoe. SimPoe employs a universal physics-based tracking controller trained on a large-scale 3D motion dataset AMASS~\cite{mahmood2019amass} as the motion corrector. Recently, video data has demonstrated its superior accessibility and scalability for learning intricate skills that are difficult or costly to capture in a laboratory environment~\cite{xie2021physics, zhang2023tennis, mao2024learning, Wang_2025_CVPR, wang2025hil}. However, a major challenge is that 3D poses from video are not physically reliable. This data therefore requires extensive post-processing or even manual correction before it can be used to train a physics-based control policy. In contrast, our method performs end-to-end learning directly from 2D pose sequences, making it applicable to 'in-the-wild' videos and adaptable to various character skeletons.
\section{Method} 
\subsection{Imitation as Reprojection Minimization}\label{sec:task_formulation}
Given a 2D motion sequence denoted by as a series of coordinates $\mathbf{X} \in \mathbb{R}^{T \times J \times 2}$, where $T$ is the motion length and $J$ is the number of keypoints (including skeleton joints or object landmarks), our goal is to learn a policy $\pi$ capable of controlling the simulated character to perform physically plausible 3D motions such that their 2D projection onto a given camera view aligns precisely with the provided 2D motion reference. Existing approaches often separately reconstruct 3D motion from the 2D evidence and learn a control policy to imitate the reconstructed 3D motion. Due to the ill-posed nature of 2D-to-3D inversion, the reconstructed 3D motions are often physically infeasible, especially in domains such as object interaction or non-human locomotion where 3D priors are unavailable. This flawed 3D supervision poses a major obstacle to motion imitation and often leads to the failure of learning. To address this, we propose to unify motion reconstruction and motion imitation by formulating a physics-based 2D motion tracking problem defined by the following:
\begin{equation}\label{eq:tracking_formulation}
\min_{\pi} \E_{s_0 \sim d(s_0)} \|P_{\pi}(C) - \mathbf{X} \| \
\quad \text{s.t.} \quad f_{\pi} = 0,
\end{equation}
where $P_{\pi}(C)$ denotes the 2D projection of the 3D joint positions synthesized by the policy $\pi$ under the camera view $C$, $d(s_0)$ is the distribution of the initial state, and $f_{\pi} = 0$ represents the physical constraints and underlying MDP dynamics of the simulated character.
The unified formulation allows for end-to-end error minimization and leverages physics constraints to regularize the resultant 3D motions, thereby ensuring physical plausibility.

\subsection{View-agnostic 2D Tracking Policy}\label{sec:policy_representation}
Optimizing solely for the reprojection objective is inherently limited by depth ambiguity, often leading to implausible or unnatural poses. To mitigate this issue, we observe that casual videos are typically captured from a variety of viewpoints, which collectively provide sufficient information to depict the motion. Building on this insight, we propose training a general 2D tracking policy designed to minimize the reprojection error for a given reference motion under arbitrary camera viewpoints. Such a policy implicitly acquires a 3D understanding of motion, as it learns to satisfy 2D reprojection constraints from diverse viewing directions present in the data. Furthermore, this cross-view generalization capability enables the policy to be effectively extended to multi-view motion tracking via feature aggregation, thereby achieving robust 3D motion tracking performance without relying on any 3D supervision.

The 2D tracking policy is represented as a neural network that receives inputs of the character's proprioceptive $\textbf{s}^{\text{prop}}_t$~\cite{peng2018deepmimic} and the observation of future 2D motion reference $\textbf{o}^{\text{2D}}_t$, then predicts the simulated character's PD (Proportional-Derivative) targets as a diagonal Gaussian distribution over the action space.

\paragraph{View-agnostic 2D Observations}
To achieve better generalization ability for arbitrary-view 2D motion tracking, we intentionally omit any explicit camera view information from the observations, thereby compelling the policy to infer the viewpoint of the 2D motion reference solely from the character's actual 2D projection within the simulation.

The proposed 2D observation is represented by:
\begin{equation}
    \mathbf{o}^{\text{2D}|C}_t = [P(\mathbf{x}_t^{\text{3D}}, C), \mathbf{x}_{t:t+L}],
\end{equation}
where $\mathbf{x}_{t:t+L} \in \mathbb{R}^{L \times J \times 2}$ is the clip of future 2D reference motion, looking ahead over $L$ frames, and $P(\mathbf{x}_t^{\text{3D}}, C) \in \mathbb{R}^{J \times 2}$ denotes the 2D projection of the 3D keypoints $\mathbf{x}_t^{\text{3D}} \in \mathbb{R}^{J \times 3}$ in simulation.

\paragraph{View Aggregation for 3D Motion Tracking}
The diverse-view data provides a well-constrained motion space, thereby leading to more physically plausible 3D motions synthesized by the view-agnostic tracking policy. However, the inherent ambiguity of a single 2D projection still persists, which limits the policy's fine-grained controllability in 3D space.  As shown in Figure~\ref{fig:overview}, we propose to adapt our policy to a multi-view tracking policy by introducing a view-aggregation technique to mitigate this issue.

We first demonstrate how to extend the capabilities of the general 2D single-view tracking policy to achieve 3D motion tracking by reframing the single-view objective into a multi-view setup. We define a multi-view tracking problem by considering $K$ multi-view 2D motions and their corresponding camera views, $\{\mathbf{X}_k, C_k\}_{k=1}^K$, which are all projected from a single underlying 3D motion sequence. 
The objective to find a policy $\pi$ that minimizes the aggregate reprojection error across all views is
\begin{equation}\label{eq:multi_view_tracking}
    \min_{\pi} \E_{s_0 \sim d(s_0)} \sum_{k=1}^K \left\| P_{\pi}(C_k) - \mathbf{X}_k \right\|
    \quad \text{s.t.} \quad f_{\pi} = 0.
\end{equation}
We hypothesize that the optimality attained by the general single-view tracking policy is sufficient to achieve optimality for the multi-view tracking problem defined above.

To enable the policy to accept 2D observations from multiple views $\{\textbf{o}^{\text{2D}|C_1}_t, \textbf{o}^{\text{2D}|C_2}_t, \cdots, \textbf{o}^{\text{2D}_{C_K}}_t\}$, we devised a view aggregation strategy for combining information across different views. Leveraging the linearity typically observed in the neural network feature space, we opt to average the feature embeddings of the 2D observations from all views. The multi-view motion tracking policy $\pi_{\text{mv}}$ adapted from the view-agnostic tracking controller is defined as:
\begin{equation}
    \pi_{\text{mv}}(a_t | \mathbf{s}^{\text{prop}}_t, \textbf{o}^{\text{2D}|C_1}_t, \cdots, \textbf{o}^{\text{2D}|{C_K}}_t) = 
    \pi(a_t | \mathbf{s}^{\text{prop}}_t, \textbf{o}^{\text{agg}}_t),
\end{equation}
where the aggregated feature $\mathbf{o}^{\text{agg}}_t$ is computed by averaging the view-specific features:
\begin{equation}
    \mathbf{o}^{\text{agg}}_t = \frac{1}{N} \sum_{i=1}^N \phi(\mathbf{o}^{\text{2D}_i}_t),
\end{equation}
where $\phi(\cdot)$ denotes the feature extractor for single-view 2D observations. The proposed view-aggregation strategy grants the view-agnostic tracking policy the capability for multi-view motion tracking without requiring any fine-tuning, thereby greatly enhancing its 3D controllability.

\subsection{Training for Single-view Tracking}\label{sec:policy_learning}
Training the view-agnostic 2D tracking policy follows a similar reinforcement learning paradigm to 3D motion tracking, with three key adaptations: 2D tracking rewards specifically designed to minimize the reprojection error, an adaptive state-initialization to resolve the lack of high-quality 3D reference poses, and an early termination strategy tailored for 2D motion imitation training.

\paragraph{2D Tracking Rewards}
To encourage the minimization of reprojection error in Eq.~\ref{eq:tracking_formulation}, we employ a distance-based reward function as the 2D tracking reward, combined with an energy consumption penalty for regularization $r_t = w_p r^p_t + w_e r^e_t$, where $w_p$ and $w_e$ are reward weights.
The 2D tracking reward $r^p_t$ is computed by measuring the discrepancy between the projected 2D keypoints and the reference 2D points:
\begin{equation}\label{eq:2d_track_reward}
    r^p_t = \exp\left(-\alpha \sum_{j=1}^J \|P(\mathbf{x}^{\text{3D}|j}_t) - \mathbf{x}^j_t\|\right),
\end{equation}
where $\mathbf{x}^j_t$ denotes the coordinates of the $j$-th reference keypoint at time $t$, $P(\mathbf{x}^{\text{3D}|j}_t)$ denotes the projected 2D coordinates of the simulated character joint, and $\alpha$ is a positive scalar. The energy consumption penalty $r^e_t$ is computed as$r^e_t = -\sum_j \left\| \dot{\textbf{q}}^j_t \cdot \tau^j_t \right\|,$ where $\dot{\textbf{q}}^j_t$ and $\tau^j_t$ denote the joint velocity and torque for joint $j$ at time $t$, respectively.

\begin{figure*}[t]
  \centering
  \includegraphics[width=1.0\textwidth]{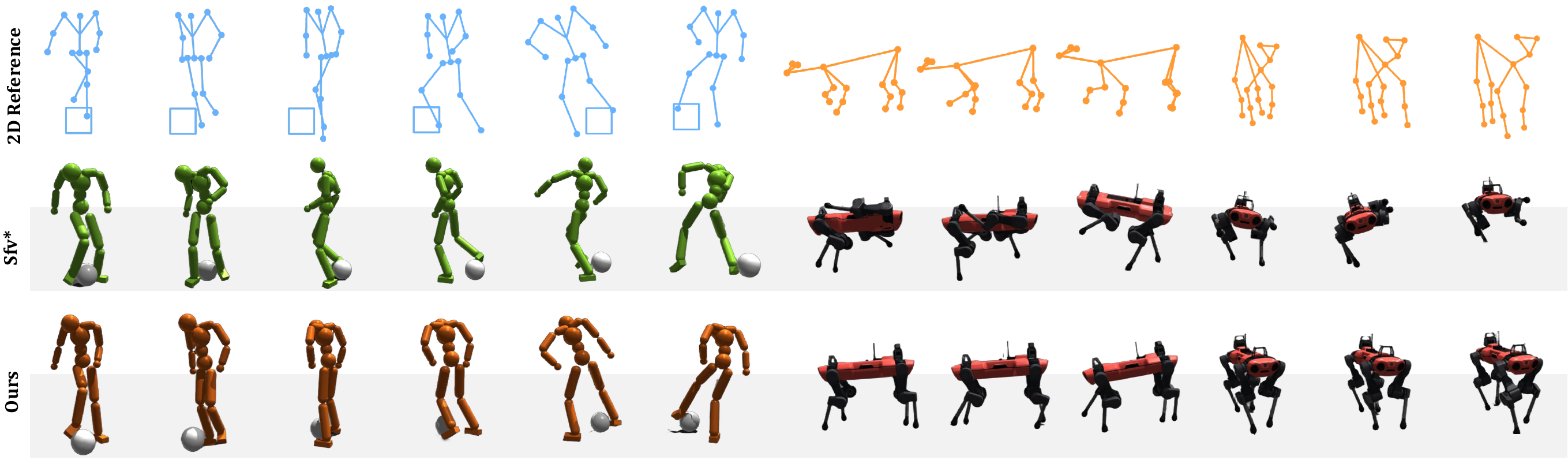}
  \caption{
  \textbf{Qualitative comparison} between our method (bottom) and the adapted Sfv*~\cite{peng2018sfv} (middle) on imitating 2D reference motions (top) from Soccer Dribble and Animal. Sfv* struggles to reproduce complex ball–foot interactions (left) and often yields unnatural motions (right) due to inaccuracies in the reconstructed 3D training data. Sfv* is adapted to learn from 3D motions reconstructed via reprojection loss minimization.
  }
  \label{fig:comparison}
\end{figure*}

\paragraph{Adaptive State Initialization}
The distribution of initial states is crucial for learning efficiency of RL. In conventional 3D motion imitation, the widely adopted Reference State Initialization (RSI)~\cite{peng2018deepmimic} relies on accurate 3D motion states, which are unavailable when only 2D data is provided. Although imperfect 3D reconstructions~\cite{yuan2021simpoe, yuan2019ego} or manually specified default poses can be utilized for state initialization, the occurrence of physically infeasible states will significantly impede the policy learning. To address this limitation, we propose to discard suboptimal initial states by evaluating the estimated tracking performance using the critic network. Inspired by the approach in \cite{peng2018sfv}, which learns a neural network policy to predict a distribution of initial states, we instead employ a data buffer to maintain a discrete representation of the initial state distribution for enhanced training stability.

Specifically, a dedicated buffer is maintained for each reference frame and is initially populated with arbitrary initial 3D poses. During training, states obtained from policy rollouts are evaluated using the critic network, and those states that achieve a high critic score are subsequently stored in the corresponding frame buffer. To facilitate prioritized exploration, the sampling probability for each stored state is set to be exponentially proportional to its score.

\paragraph{Reprojection-based Early Termination}
To ensure training efficiency and avoid unrecoverable states of the simulated character, we introduce an early termination mechanism based on the reprojection error. Specifically, an episode is terminated when the character's projected pose deviates significantly from the reference 2D pose.

\subsection{Hierarchical Control via 2D Interface}\label{sec:hierachical_control}
The trained view-agnostic 2D tracking policy can be extended to generative tasks via integration with a kinematic 2D motion generative model, forming a hierarchical controller. To enable real-time, infinite-length motion synthesis, we propose a transformer-based autoregressive 2D motion generator. This GPT-like architecture ensures on-the-fly generation while preserving high-quality motion characteristics. To efficiently encode the global movements in 2D motion, we introduce a canonical 2D motion representation, facilitating the learning of our 2D motion generator.

\paragraph{Canonical Representation of 2D Motion}
Raw 2D keypoint coordinates often exhibit significant variance due to global character movement, such as lateral displacement (e.g., moving from right to left) or changes in scale caused by translation along the camera axis (e.g., moving closer to or farther from the camera). This high variance makes training the generative model challenging. To mitigate this problem, we propose a canonicalized 2D motion representation that is more efficient for neural network learning while still preserving the necessary global motion information.

More specifically, we compute the root translation $x^{\text{root}}$, a scale factor $s$, and the local pose $\bar{x}$ using the formulation $\bar{x} = (x - x^{\text{root}}) / s$ for each frame. The relative scale change between consecutive frames is given by $\delta s_t = \log (s_t / s_{t-1})$, while the normalized shift in root translation is defined as $\delta x_t^{\text{root}} = (x_t^{\text{root}} - x_{t-1}^{\text{root}}) / s_t$. The relative 2D motion sequence is then represented as $\textbf{x}^{\text{can}} = [x_0^{\text{can}}, x_1^{\text{can}}, ..., x_T^{\text{can}}]$, where each $x_t^{\text{can}} = (\bar{\text{x}}_t, \delta x_t^{\text{root}}, \delta s_t)$. Given the scale and root translation of the initial frame, $\text{x}_0^{\text{root}}$ and $s_0$, we can convert between the absolute and relative 2D motion representations. The forward and inverse transformations are defined as:
\begin{equation}
    \textbf{X}^{\text{can}} = G(\textbf{X}), \quad \textbf{X} = G^{-1}(\textbf{X}^{\text{can}}, x_0^{\text{root}}, s_0).
\end{equation}

\paragraph{2D Motion Tokenizer}
We employ a Vector-Quantized Variational Autoencoder (VQ-VAE)~\cite{van2017neural} to compress redundant information and discretize the canonical 2D motion sequence into a sequence of compact tokens. To maintain accurate global translation and scale consistency over long sequences, we introduce an additional absolute 2D reconstruction term into the loss function, computed by transforming the canonical representation back to the global coordinate space:
\begin{equation}
    \mathcal{L}_{\text{rec}} = \|\mathbf{X}^{\text{can}} - \hat{\mathbf{X}}^{\text{can}}\| + \omega \|G^{-1}(\mathbf{X}^{\text{can}}) - G^{-1}(\hat{\mathbf{X}}^{\text{can}})\|,
\end{equation}
where $\omega$ is a weighting factor for the absolute loss term, and $\hat{\mathbf{x}}^{\text{can}}$ denotes reconstructed canonical 2D motion by the VQ-VAE. Furthermore, to achieve autoregressive token detokenization, we adopt causal convolutional layers~\cite{nauta2019causal} in the VQ-VAE architecture.

\begin{table}
\centering
\resizebox{\linewidth}{!}{
\begin{tabular}{lccccccc} 
\toprule
\multirow{2}{*}{Method} & 
\multicolumn{4}{c}{Soccer Dribble} & 
\multicolumn{3}{c}{Animal} \\
\cmidrule(lr){2-5}\cmidrule(lr){6-8}
& Succ.$^\uparrow$ & $E_{2D}$$^\downarrow$ & $E_{O2D}$$^\downarrow$ & Jitters$^\downarrow$ & Succ.$^\uparrow$ & $E_{2D}$$^\downarrow$ & Jitters$^\downarrow$\\
\midrule
Sfv*~\cite{peng2018sfv} & 47.8 & 19.9 & 38.2 & 2.62 & 50.0 & 68.9 & 9.20 \\
\midrule
Ours & \textbf{91.3} & \textbf{17.1} & \textbf{17.5} & \textbf{1.69} & \textbf{83.3} & \textbf{26.8} & \textbf{3.36} \\
\bottomrule
\end{tabular}}
\caption{
\textbf{Quantitative comparison} of our method against the two-stage pipeline Sfv*~\cite{peng2018sfv} on our collected Soccer Dribble and Animal datasets sourced from in-the-wild videos. Sfv* is adapted to learn from 3D motions reconstructed via reprojection loss minimization.}
\label{table:quat_comparison}
\end{table}

\paragraph{Autoregressive Generator}
By tokenizing the 2D motion sequence into discrete codebook indices using VQ-VAE, $\mathbf{c} = [c_0, c_1, \dots, c_{T/l}]$, the data distribution is modeled as a discrete, factorized distribution: $p(\mathbf{c}) = \prod_{i=1}^{T/l} p(c_i \mid c_0, ..., c_{i-1})$ where $c_i \in \{1, ..., K\}$. We employ a causal transformer to represent this distribution, denoted as $p_{2D}(c_0, c_1, ..., c_{T/l} \mid y)$, where $y$ is an optional conditioning variable (with $y=\emptyset$ for unconditional generation). The model is trained by minimizing the cross-entropy loss between the target token index and the predicted probability. 

\paragraph{Interface for Hierarchical Control}
The trained view-agnostic 2D motion tracking policy can be integrated with our 2D motion generators, forming a hierarchical controller capable of a wide range of generative tasks. The generated $L$ 2D motion reference frames $\hat{\mathbf{x}}_{t:t+L}^{\text{can}}$, are first converted to the global representation via $\hat{\mathbf{x}}_{t:t+L} = G^{-1}(\hat{\mathbf{x}}_{t:t+L}^{\text{can}})$. We regard the generated 2D motions in the global coordinate as the reference motion for our tracking policy. Benefiting from our view-agnostic training, the viewpoint used for projection can be arbitrarily selected. This design eliminates rigid constraints on viewpoints, enhancing the controller’s applicability across diverse scenarios.
\begin{table*}
\centering
\resizebox{\textwidth}{!}{
\begin{tabular}{cccccccccccccc} 
\toprule
\multirow{2}{*}{Training Data} & \multirow{2}{*}{Inputs} & 
\multicolumn{4}{c}{AIST++~\cite{li2021ai}-Train} & 
\multicolumn{4}{c}{AIST++~\cite{li2021ai}-Test} & 
\multicolumn{4}{c}{2D Motion Generator} \\
\cmidrule(lr){3-6}\cmidrule(lr){7-10}\cmidrule(lr){11-14}
& & Succ.$^\uparrow$ & $E_{3D}$$^\downarrow$ & $E_{2D}$$^\downarrow$ & Jitters$^\downarrow$ & Succ.$^\uparrow$ & $E_{3D}$$^\downarrow$ & $E_{2D}$$^\downarrow$ & Jitters$^\downarrow$ & Succ.$^\uparrow$ & $E_{2D}$$^\downarrow$ & FID$^\downarrow$ & Jitters$^\downarrow$ \\
\midrule
3D & 3D & 92.1 & 114.5 & 18.9 & 2.81 & 89 & 141 & 21.3 & 2.99 & - & - & - & - \\
\midrule
\multirow{3}{*}{2D} & 1-view 2D & 88.4 & 220.0 & 35.1 & 1.64 & 82.5 & 254.6 & 38.6 & 1.79 & 86.9 & 42.1 & 1.52 & 2.08 \\
& 2-view 2D & 90.1 & 147.8 & 22.1 & 1.58 & 88.0 & 164.9 & 24.5 & 1.60 & 92.1 & 28.9 & 2.44 & 2.07 \\
& 3-view 2D & 90.4 & 144.3 & 21.9 & 1.58 & 88.9 & 161.5 & 24.1 & 1.60 & 88.6 & 33.1 & 2.84 & 1.90 \\
\bottomrule
\end{tabular}}
\caption{
\textbf{Quantitative evaluation} of large-scale motion imitation on AIST++~\cite{li2021ai} and on 2D motions generated by our autoregressive 2D motion generator. Our approach achieves performance comparable to baselines trained on ground-truth 3D data.
}
\label{table:quat_tracking2}
\end{table*}
\begin{table}
\centering
\resizebox{0.85\linewidth}{!}{
\begin{tabular}{ll ccc} 
\toprule
Motion & Method & FID$^\downarrow$ & Succ.$^\uparrow$ & Jitters$^\downarrow$ \\
\midrule
\multirow{3}{*}{Soccer Dribble} & Diffusion & 6.29 & 0.13 & 2.16 \\
& AR (Ours) & \textbf{5.16} & \textbf{0.71} & \textbf{1.91} \\
\cmidrule(lr){1-5}
\multirow{3}{*}{AIST++} & Diffusion & 5.92 & 0.44 & \textbf{1.87} \\
& AR (Ours) & \textbf{2.44} & \textbf{0.92} & 2.07 \\
\bottomrule
\end{tabular}}
\caption{
\textbf{Comparative evaluation of generative models} within the hierarchical control framework on the Soccer Dribble and AIST++ datasets.
}
\vspace{-0.4cm}
\label{table:gen_method_comparison}
\end{table}
\begin{figure*}[t]
    \centering
    \includegraphics[width=1.0\textwidth]{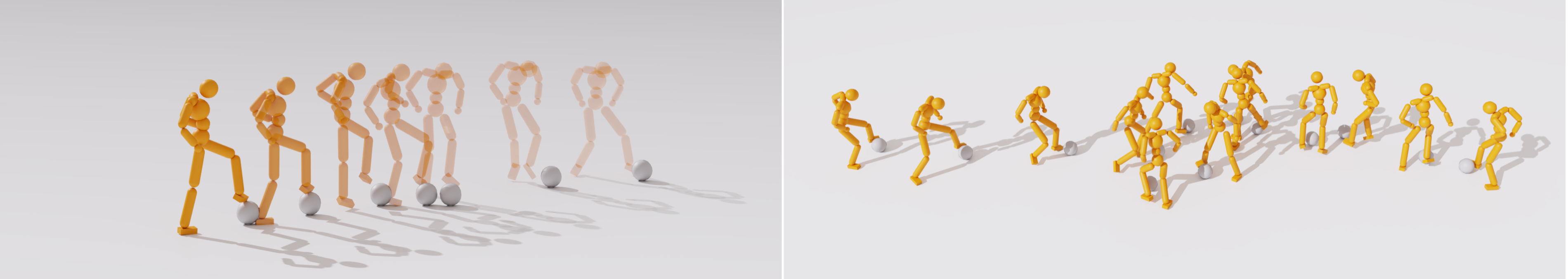}
    \caption{
    \textbf{Visualization of synthesized ball-dribbling motions} using our hierarchical controller. Our framework learns diverse skills from 2D motion data, exhibits natural transitions between different dribbling styles (left), and demonstrates seamless composition and switching between directional movements.
    }
    \label{fig:dribble}
    \vspace{-5pt}
\end{figure*}

\section{Experiments}
\subsection{Experimental Setup}
\paragraph{Data}
We evaluate the effectiveness of our framework using both in-the-wild videos and public datasets. To assess its ability to learn from challenging scenarios with scarce 3D data, such as HOI and non-human motions, we curate two new datasets of soccer dribbling and animal motions. \textbf{Soccer Dribble} is collected from online videos showcasing a diverse range of complex soccer dribbling skills that involve intricate football interaction dynamics. We apply ViTPose~\cite{xu2022vitpose} to detect and extract 2D human poses and ball bounding boxes from these videos. \textbf{Animal} is also sourced from online videos and includes various dog movements such as walking, running, and jumping. The 2D keypoints are detected and processed using the same pipeline as the Dribble dataset. For benchmarking, we also include the \textbf{AIST++} dataset~\cite{li2021ai}, a large-scale public dataset containing dynamic dance videos captured from multiple viewpoints with annotated 3D human poses. Following the protocol in~\cite{li2024lifting}, we generate a single-view 2D motion data by randomly selecting a single camera view per sequence.

\paragraph{Metrics}
We adopt the following metrics for quantitative evaluation. \textit{Success Rate} measures the robustness of the control policy and is defined as the percentage of successfully tracked motions over all reference motions. A trial is considered unsuccessful if the maximum reprojection error exceeds 100 pixels. \textit{2D Tracking Error} evaluates the policy’s tracking accuracy in 2D space and is computed as the average reprojection error between the character’s projected poses and the reference motion. \textit{2D Object Tracking Error} is similarly defined but measures the accuracy of object tracking in human–object interaction (HOI) tasks. \textit{3D Tracking Error} assesses the tracking accuracy with respect to 3D reference motions by averaging the positional joint errors across all frames. \textit{Jitters} quantifies the physical plausibility of the synthesized motion by computing the third-order derivatives of joint positions as a measure of motion smoothness.
Finally, \textit{FID} measures the distributional difference between the synthesized and reference motions. Following the protocol in~\cite{kapon2024mas}, FID scores are computed on the 2D projections of the synthesized poses.

\paragraph{Implementation Details} 
The control policy is running at 30Hz and interacts with the physics simulator Isaac Gym~\cite{makoviychuk2021isaac}, which simulates dynamics at 60Hz. The control policy is an MLP network with 512, 256, 256 hidden units. For the Vector-Quantized Variational Autoencoder (VQ-VAE), we set the codebook size to 512 with an embedding dimension of 128. The autoregressive transformer employs a decoder architecture with four layers, four attention heads, and an embedding dimension of 128. We simply optimize the reprojection loss obtain the initial poses as initial states for single-view tracking training. Note that our approach is not sensitive to the initial states, we can obtain them through various methods, including state-of-the-art pose estimation techniques or manual specification. The training of the single-view tracking policy takes approximately one week for both AIST++ motions and Dribble motions, using four NVIDIA P40 GPUs. The imitation of animal motions takes about three days with the same resources.
 
\subsection{Evaluation on Motion Imitation}
\paragraph{Comparison with Two-stage Approach}
We evaluate our approach on the challenging HOI and non-human motion datasets (Soccer Dribble and Animal). We also conduct a detailed comparison against the typical two-stage baseline, Sfv~\cite{peng2018sfv}, which first estimates 3D poses before imitating with the reconstructed trajectories. Since the original Sfv~\cite{peng2018sfv} was not designed for domains like HOI or animal motions, we adapted its motion reconstruction stage for a fair comparison. For the Soccer Dribble dataset, we utilized SLAHMR~\cite{ye2023slahmr} to obtain the 3D human poses. The ball's 3D position was subsequently reconstructed by minimizing the reprojection loss on its bounding box and incorporating the estimated depth of the ball. For the Animal motion data, we simply solve an inverse kinematics problem by optimizing the reprojection loss since there is no data prior available.

The quantitative comparison results, presented in Table~\ref{table:quat_comparison}, demonstrate the superior performance of our approach against the baseline. Our method achieves both a higher learning success rate and lower motion jitters and with lower 2D tracking error and object 2D tracking error. The qualitative comparison (Figure~\ref{fig:comparison}) illustrates the limitation of the baseline: Sfv* fails to learn complex ball turnaround interactions and exhibits implausible quadruped locomotion due to imperfect training motions. In contrast, our approach successfully synthesizes realistic motions that accurately align with the 2D references.

\vspace{-0.1cm}
\paragraph{General Motion Tracking}
The view-agnostic tracking policy is generalizable and can achieve zero-shot multi-view motion tracking via view-aggregation. To thoroughly evaluate its tracking ability, we train our approach on the AIST++ dataset and compare it with the 3D motion imitation baseline trained using ground-truth 3D motions.

As shown in Table~\ref{table:quat_tracking2}, the trained single-view tracking policy successfully generalizes to unseen test and generated motions. Furthermore, the proposed view-aggregation mechanism grants the policy multi-view tracking capability, boosting 3D motion understanding with lower 3D motion tracking error, and even achieving performance comparable to the 3D baseline trained on ground-truth data.

\subsection{Evaluation on Hierarchical Control}
To evaluate the effectiveness of our hierarchical control framework using 2D motion interfaces, we train the 2D motion generator on the AIST++ and Soccer Dribble datasets. Qualitatively, the trained motion generator successfully demonstrates the ability to synthesize complex trajectories by seamlessly combining different motion skills, as illustrated by the transition between various dribbling skills in Figure~\ref{fig:dribble}. We further benchmarked our proposed autoregressive(AR) 2D motion generator against the diffusion-based model baseline~\cite{tevet2024closd}. As detailed in Table~\ref{table:gen_method_comparison}, our AR generator demonstrates superior performance in synthesizing more realistic 2D guidance, achieving a lower FID score and higher success rate compared to the baseline.

\subsection{Ablation Study}
\begin{figure}[t]
    \centering
        \includegraphics[width=1.0\linewidth]{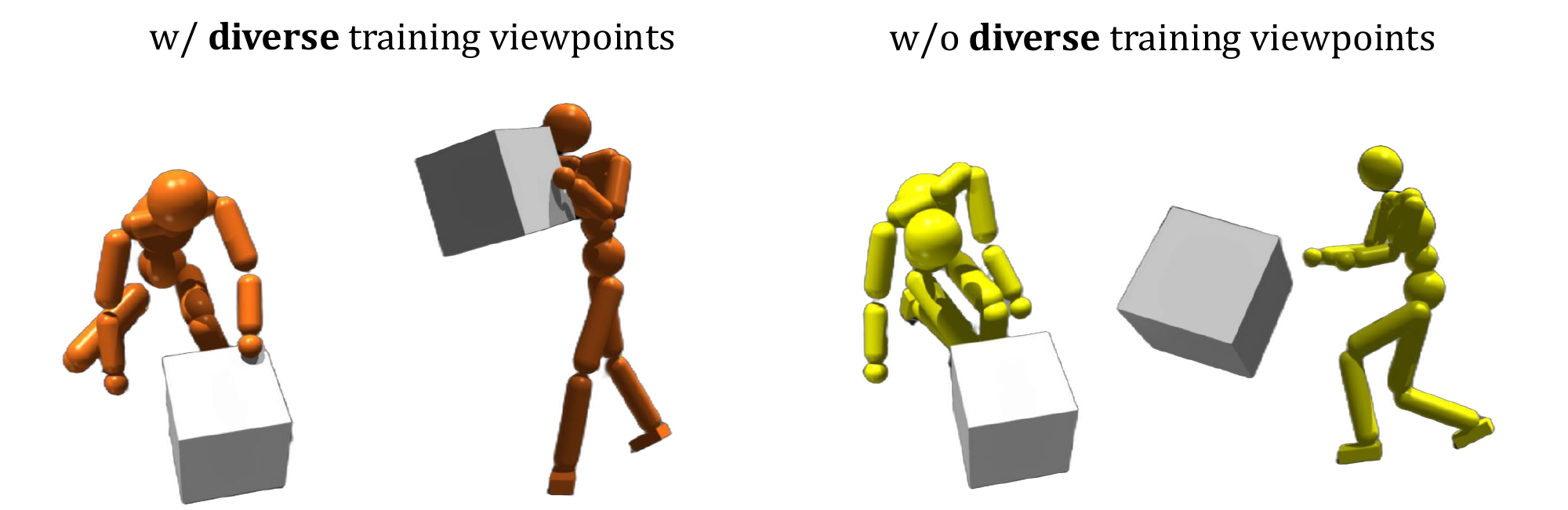}
    \caption{
    \textbf{Exploiting viewpoint diversity in the data.}
    Our view-agnostic tracking policy learns a stronger 3D understanding of character poses when trained on diverse viewpoints (left). In contrast, a policy trained on homogeneous viewpoints exhibits unnatural behaviors and fails to acquire interaction skills (right).
    }
    \vspace{-0.5cm}
    \label{fig:ablation_view_diversity}
\end{figure}

\paragraph{View Diversity in Dataset}
To investigate how effectively our approach benefits from the diversity of viewpoints in the training data, we conducted an ablation study. We trained two variants of the policy: one using 2D motions projected from the same 3D instance with diverse viewpoints, and a second using 2D motions with homogeneous viewpoints. As illustrated in Figure~\ref{fig:ablation_view_diversity}, the policy trained on diverse-view motions synthesizes significantly more natural-looking behavior than that trained solely on homogeneous-view motions. Notably, the homogeneous-view policy failed to execute the challenging ``lift the box" interaction, underscoring the importance of view diversity for learning complex skills.

\begin{figure}[t]
    \centering
    \includegraphics[width=1.0\linewidth]{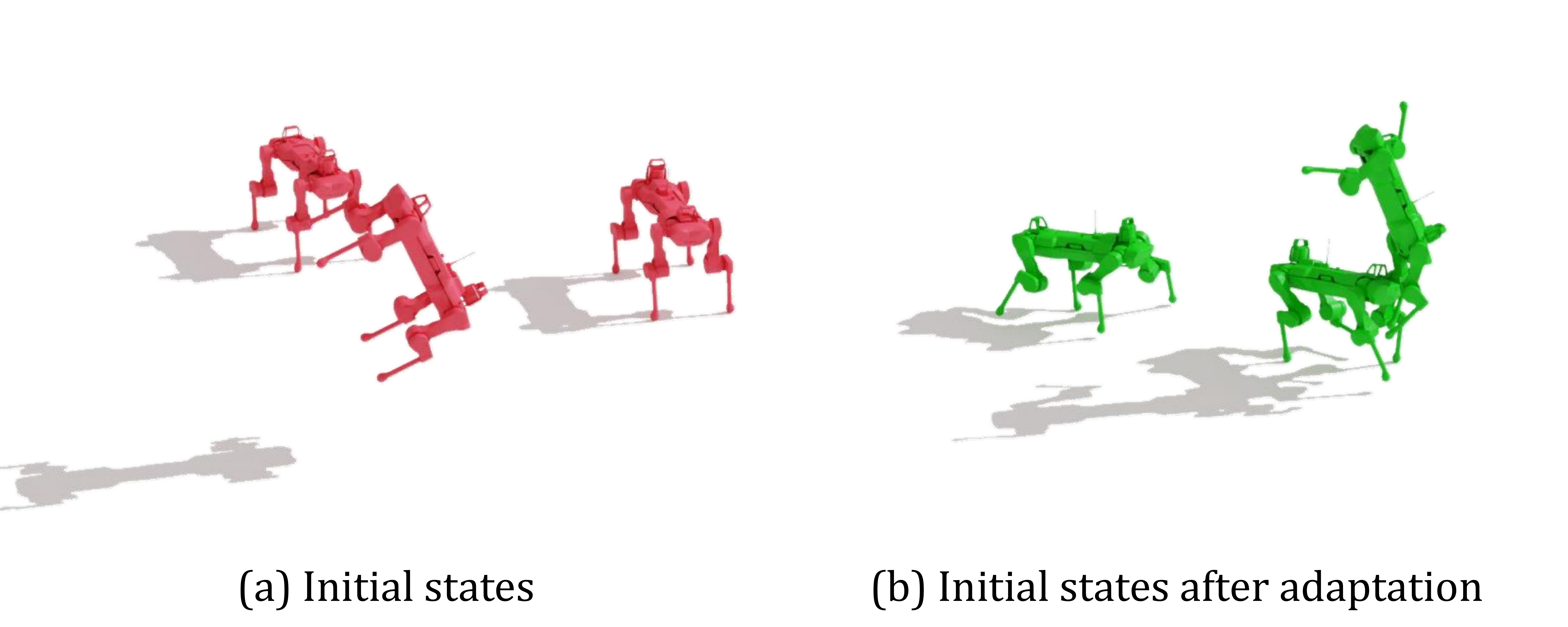}
    \caption{
    \textbf{Effectiveness of adaptive state initialization.}
    Our method adaptively updates and refines physically implausible initial states (left) into more feasible ones (right), enabling more efficient policy training.
    }
    \vspace{-0.5cm}
    \label{fig:init_states}
\end{figure}

\paragraph{Adaptive State Initialization}
We evaluate the effectiveness of our proposed adaptive state initialization strategy for single-view tracking training. As shown in Figure~\ref{fig:init_states}, the adaptive state initialization can significantly update the initial states from the erroneous initialization.

\vspace{-0.2cm}
\section{Conclusion}
\vspace{-0.1cm}
In this work, we present a novel approach to leveraging video data for learning physics-based character controllers by training a view-agnostic tracking policy to directly imitate 2D motion sequences extracted from videos. Our framework generalizes robustly across diverse motions and character articulations, successfully learning complex HOI motions and agile non-human movements without access to any 3D motion data. Furthermore, by incorporating view aggregation, our trained policy achieves 3D tracking performance comparable to baselines trained directly on ground-truth 3D data. Finally, by integrating the policy with a proposed autoregressive 2D motion generator, we establish a hierarchical controller capable of generating physically plausible motions, demonstrating significant potential for a variety of downstream tasks.
{
    \small
    \bibliographystyle{ieeenat_fullname}
    \bibliography{reference}
}
\clearpage
\setcounter{page}{1}
\maketitlesupplementary
\appendix

\section{Preliminary of Reinforcement Learning}
We approach the single-view 2D motion tracking using reinforcement learning, which formulates a Markov Decision Process (MDP) defined by the tuple $\mathcal{M} = (\mathcal{S}, \mathcal{A}, \mathcal{T}, \mathcal{R})$, where $\mathcal{S}$ represents states, $\mathcal{A}$ actions, $\mathcal{T}$ environment dynamics, and $\mathcal{R}$ reward functions. The tracking controller is modeled as a policy $\pi(a_t|s_t)$, which samples actions $\textbf{a}_t \in \mathcal{A}$ based on the current state $\textbf{s}_t \in \mathcal{S}$. We collect state-action-reward trajectories $\tau = \{(\textbf{s}_t, \textbf{a}_t, r_t)\}_{t=1}^T$ in the simulator, following the policy’s actions, the simulation dynamics $\mathcal{T}(\textbf{s}_{t+1}|\textbf{s}_t, \textbf{a}_t)$, and the reward function $r_t = \mathcal{R}(\textbf{s}_t, \textbf{a}_t)$. The training objective is to learn an optimal policy $\pi^*$ that maximizes the expected cumulative return $R = \mathbb{E}_\pi\left[\sum_t \gamma^t r_t\right]$ over the collected trajectories. 

\section{Implementation Details}
\subsection{Data Preprocessing}
We adopt a pinhole camera model for reprojection computation, where each camera view $C$ is parameterized by $ C = (R_C, \tau_C, f_C)$, with $R_C$ and  $\tau_C$  denoting the extrinsic rotation and translation, and $f_C$ representing the intrinsic parameters.

For in-the-wild videos, we estimate the camera parameters \(C\) and the initial states \(s_0\) directly from the extracted 2D motion sequences. We jointly optimize the camera extrinsics and the 3D character poses by minimizing a reprojection loss, while keeping the intrinsics fixed. The resulting 3D poses are then used as the initial states for single-view tracking training. To avoid ground penetration, we include an additional regularization term that penalizes any body points falling below the ground plane.

\subsection{Proprioceptive States Representation}
The proprioceptive states $o_\text{prop}$ consists of the following elements: root height $\textbf{h}_t \in \mathbb{R}$, root velocity in the local coordinate frame $\textbf{v}_t \in \mathbb{R}^3$, root angular velocity in the local frame $\textbf{w}_t \in \mathbb{R}^3$, joint rotations in local joint coordinates $\textbf{q}_t \in \mathbb{R}^{6 \times J}$, joint velocities in local joint coordinates $\dot{\textbf{q}}_t \in \mathbb{R}^{3 \times J}$, and the Cartesian positions of key body links in the local root frame $\textbf{p}_t \in \mathbb{R}^K$.

\subsection{Algorithm of Adaptive State Initialization}
The details of the proposed adaptive state initialization strategy are provided in Algorithm \ref{alg:adaptive_state_buffer}.
\begin{algorithm}
\caption{Initial State Distribution Update and Sampling}
\label{alg:adaptive_state_buffer}
\begin{algorithmic}[1]
\State \textbf{Input:} Initial state distribution $d(s_0)$, Reference motion frames $\mathbf{X} = \{\mathbf{x}_0, \dots, \mathbf{x}_T\}$, Critic network $V_\phi$
\State \textbf{Output:} Updated initial state buffer $\mathcal{B}$
\State \textbf{Initialize:} Create state buffer $\mathcal{B} = \{ \mathcal{B}_t \}_{t=0}^T$, where each $\mathcal{B}_t$ is dedicated to frame $\mathbf{x}_t$.
\State \hrulefill

\Function{SampleInitialState}{$\mathcal{B}_t$}
    \State $\text{Scores} \gets \{ \text{Score}(s) \mid s \in \mathcal{B}_t \}$
    \State $\text{Probabilities} \gets \text{Normalize}(\exp(\beta \cdot \text{Scores})) $ \Comment{$\beta$ controls prioritization}
    \State \textbf{return} $\text{Sample}(\mathcal{B}_t, \text{Probabilities})$
\EndFunction

\State \hrulefill

\Function{UpdateStateBuffer}{$\mathcal{B}$, $\text{Rollouts}$}
    \For{each state $\{s_t, \mathbf{x}_t\}$ at frame $\mathbf{x}_t$ in $\text{Rollouts}$}
        \State $\text{CriticScore} \gets V_\phi(s_t)$ \Comment{Estimate value/tracking performance}
        \If{$\text{CriticScore} > \text{Threshold}$ or $|\mathcal{B}_t| < \text{MinSize}$}
            \State $\text{Score}(s_t) \gets \text{CriticScore}$
            \State $\mathcal{B}_t \gets \mathcal{B}_t \cup \{s_t\}$ \Comment{Store state in the buffer for frame $t$}
            \If{$|\mathcal{B}_t| > \text{MaxSize}$}
                \State $\mathcal{B}_t \gets \text{RemoveLowestScore}(s', \mathcal{B}_t)$ \Comment{Maintain buffer size limit}
            \EndIf
        \EndIf
    \EndFor
\EndFunction

\end{algorithmic}
\end{algorithm}


\subsection{2D Motion Tokenizer}
We employ a VQ-VAE as the motion tokenizer.
The VQ-VAE comprises an encoder $E(\cdot)$ and a decoder $D(\cdot)$. The encoder converts the relative 2D motion sequence $\textbf{x}^{\text{rel}}$ into a sequence of lower-dimensional latent vectors, $\textbf{z} = [z_0, z_1, ..., z_{T/l}]$, where $l$ denotes the temporal downsampling factor. Each latent vector $z_i$ is then quantized to its nearest entry in a learned codebook of discrete embeddings $\textbf{C} = \{c_k\}_{k=1}^K$ by $\hat{z}_i = \underset{c_k \in \mathcal{C}}{\arg\min} \; \| z_i - c_k \|_2.$

The decoder reconstructs the original relative 2D motions from the quantized latent representations, i.e., $\hat{\mathbf{x}}^{\text{rel}} = D(\hat{\mathbf{z}})$, where $D(\cdot)$ denotes the decoder and $\hat{\mathbf{z}}$ are the quantized latents.

The VQ-VAE is trained by minimizing the following loss function:
\begin{equation}
\mathcal{L}_{\text{vq}} = \mathcal{L}_{\text{recon}} + 
\| \mathrm{sg}[\mathbf{Z}] - \hat{\mathbf{Z}} \|_2^2
+ \beta \| \mathbf{Z} - \mathrm{sg}[\hat{\mathbf{Z}}] \|_2^2,
\end{equation}
where $\mathcal{L}_{\text{recon}}$ is the reconstruction loss, the second term encourages the embedding vectors to be close to the encoder outputs, and the third term penalizes the encoder for deviating from the quantized embeddings. Here, $\mathrm{sg}[\cdot]$ denotes the stop-gradient operator, and $\beta$ is a hyperparameter that balances the commitment loss. We stabilize training through exponential moving average (EMA) updates on codebook embeddings~\cite{zhang2023generating}.

\subsection{Autoregressive Generator}
We use a Transformer decoder architecture for the autoregressive generator to predict motion tokens. Each discrete token index is first mapped into an embedding, and a conditional embedding, which can be derived from the given control input or set to zero for unconditional generation, is prepended to the sequence. Absolute positional embeddings are added to encode the temporal order.

For generative tasks that are more sensitive to depth ambiguity in single-view 2D inputs, such as dance motion generation, we train a multi-view 2D generator to supply an additional viewpoint, helping to resolve missing depth cues.
To construct training data for the multi-view generator, we collect 3D motion states from the simulated character while it tracks the original single-view 2D sequences, then project them into multiple 2D views to obtain paired multi-view motion samples.
We represent the resulting multi-view 2D motions as sequential data, where each frame contains multiple motion tokens ordered by view ID.

\subsection{Key Hyperparameters}
\label{app:param}

    \begin{table}[tbp!]
        \centering
        \caption{Key Hyperparameters for the Policy Learning (PPO).}
        \begin{tabular}{ll}
        \toprule 
        \textbf{Hyperparameter} & \textbf{Value} \\
        \midrule 
        Optimizer & Adam \\
        Batch Size & 16,384 \\
        Learning rate & $2 \times 10^{-5}$ \\
        Total Training Frames & $5 \times 10^{10}$ \\
        Number of Environments & 16,384 \\
        Rollout Length & 32 \\
        PPO Epochs ($N_{\text{epochs}}$) & 3 \\
        Value Loss Coefficient & 5  \\ 
        Discount Factor $\gamma$ & 0.99  \\ 
        GAE $\tau$ & 0.95  \\ 
        PPO clipping & 0.2  \\ 
        \bottomrule
        \end{tabular}
        \label{table:app_params}
    \end{table}

    The key hyperparameters for policy learning are listed in Table~\ref{table:app_params}. Our PPO implementation is based on RLGames~\cite{rl-games2021}. The hyperparameters for the VQ-VAE and the autoregressive generator are provided in Table~\ref{table:vqvae_params} and Table~\ref{table:ar_gen_params}, respectively.

    \begin{table}[tbp!]
    \centering
    \caption{Key Hyperparameters for the VQ-VAE Motion Tokenizer.}
    \begin{tabular}{lc}
    \toprule
    \textbf{Hyperparameter} & \textbf{Value} \\
    \midrule
    \multicolumn{2}{l}{\textbf{Training Configuration}} \\
    Optimizer & Adam \\
    Learning Rate & $5 \times 10^{-5}$ \\
    Batch Size & 128 \\
    Training Steps & 600,000 \\
    \midrule
    \multicolumn{2}{l}{\textbf{Codebook Parameters}} \\
    Codebook Size ($K$) & 512 \\
    Embedding Dimension ($D_{\text{embed}}$) & 128 \\
    \midrule
    \multicolumn{2}{l}{\textbf{Architecture (ResNet Encoder/Decoder)}} \\
    Model Width (Channels) & 512 \\
    Number of ResNet Blocks & 2 \\
    Temporal Downsampling Factor & 2 \\
    \bottomrule
    \end{tabular}
    \label{table:vqvae_params}
\end{table}

\begin{table}[tbp!]
    \centering
    \caption{Key Hyperparameters for the Autoregressive Generator.}
    \begin{tabular}{lc}
    \toprule
    \textbf{Hyperparameter} & \textbf{Value} \\
    \midrule
    \multicolumn{2}{l}{\textbf{Training Configuration}} \\
    Optimizer & Adam \\
    Learning Rate & $5 \times 10^{-5}$ \\
    Batch Size & 128 \\
    Training Steps & 600,000 \\
    \midrule
    \multicolumn{2}{l}{\textbf{Architecture (Transformer Decoder)}} \\
    Transformer Layers ($N_{\text{layers}}$) & 4 \\
    Attention Heads ($N_{\text{head}}$) & 4 \\
    Embedding Dimension & 256 \\
    \midrule
    \multicolumn{2}{l}{\textbf{Data \& Sequence Parameters}} \\
    Motion Clip Length (Frames) & 60 \\
    \bottomrule
    \end{tabular}
    \label{table:ar_gen_params}
\end{table}

\section{Additional Results}

\begin{figure}[t]
    \centering
    \includegraphics[width=1.0\linewidth]{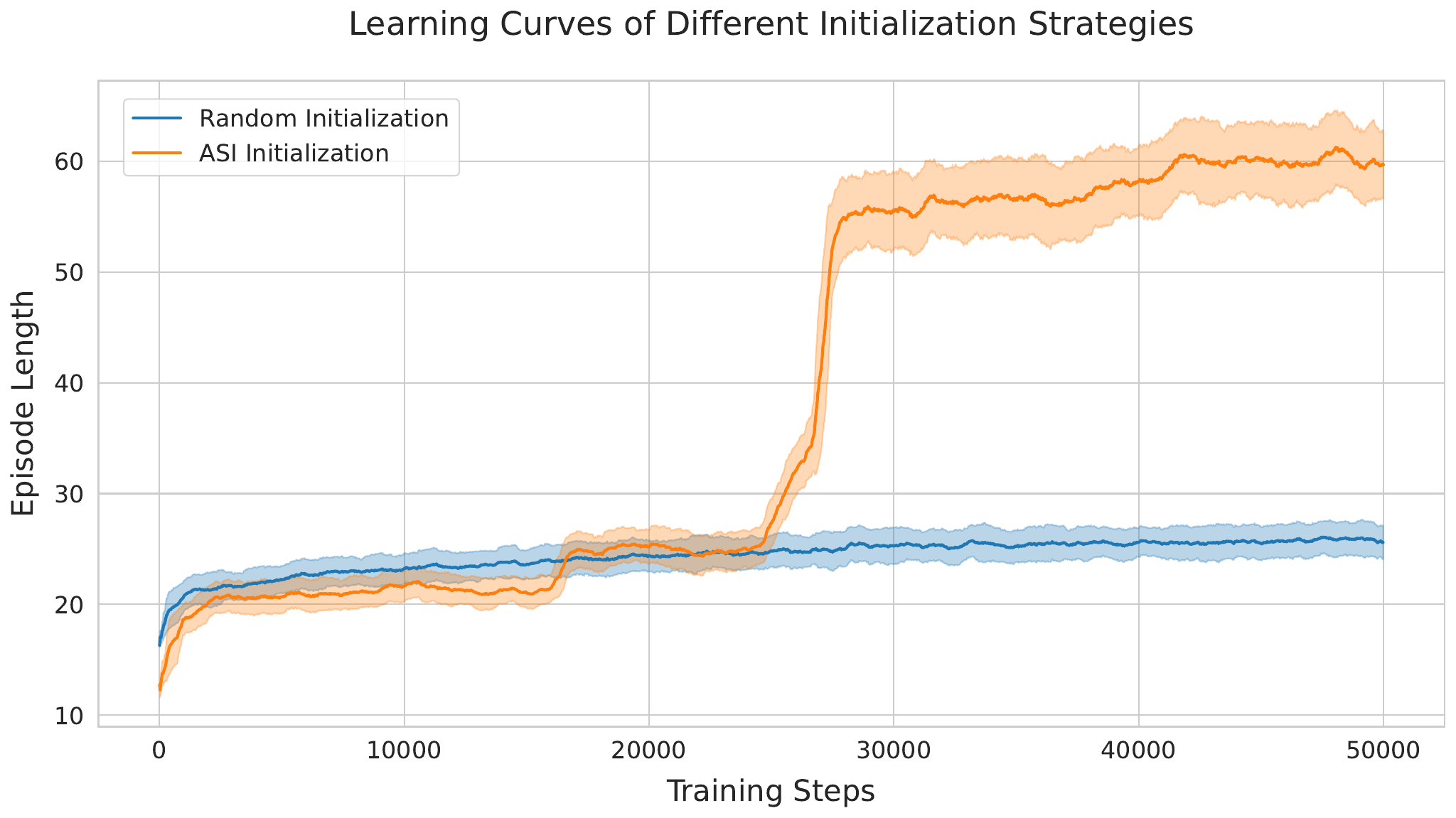}
    \caption{
    \textbf{Learning curve of 2D motion imitation with and without adaptive state initialization.} Adaptive optimization of the initial state distribution leads to significantly faster convergence in 2D motion imitation.}
    \label{fig:learning_curve}
\end{figure}

\begin{figure}[t]
    \centering
    \includegraphics[width=0.8\linewidth]{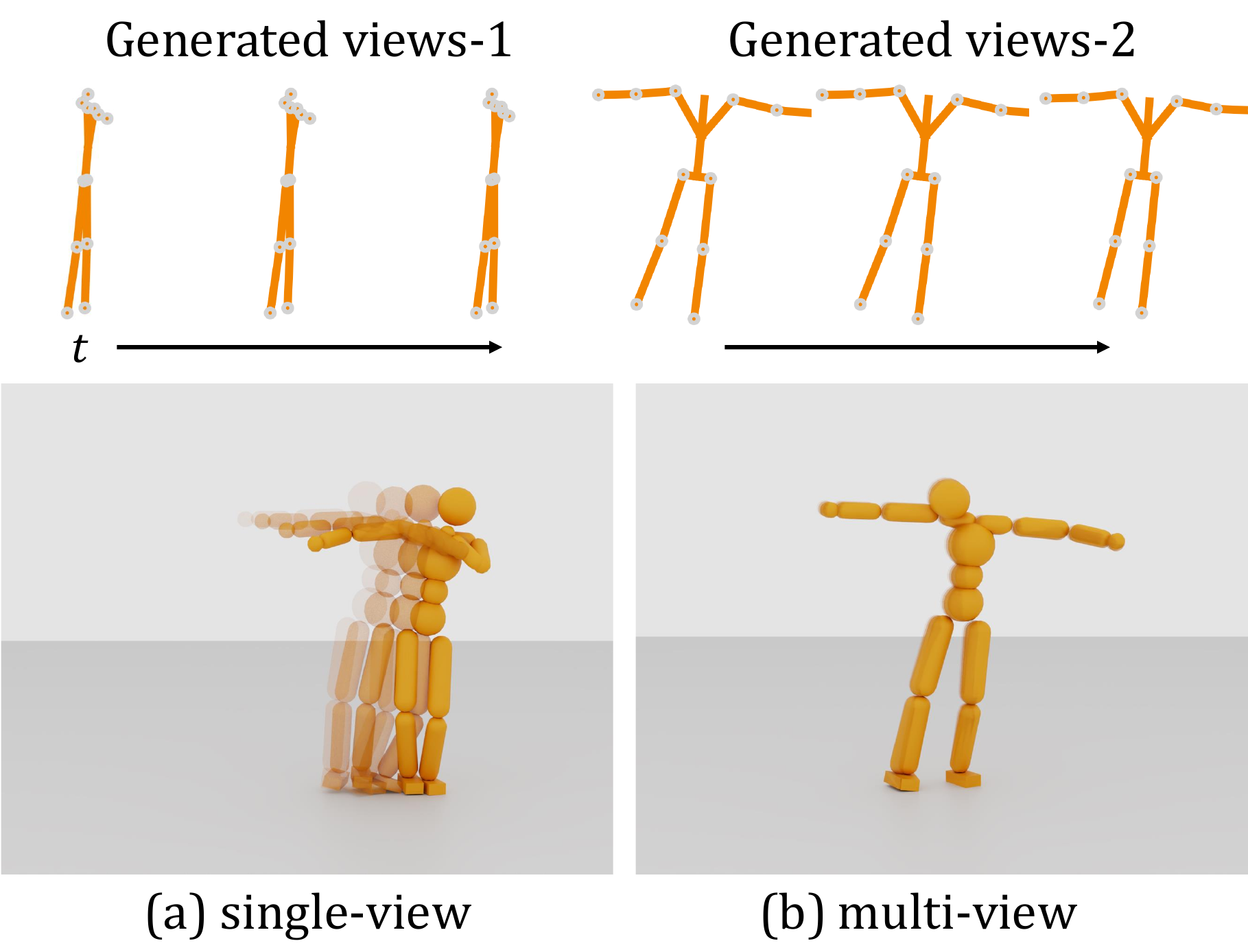}
    \caption{\textbf{Impact of multi-view guidance on resolving depth ambiguity.} Top: Reference keyframes generated from two viewpoints. Bottom: (a) Tracking guided only by View-1 results in instability and sliding due to depth ambiguity. (b) Tracking guided by both views successfully resolves these ambiguities, allowing the controller to recover stable and physically plausible 3D poses.}
    \label{fig:ambiguity_dance}
\end{figure}

\paragraph{State Initialization Strategy}
We compare the learning curves of average tracking duration for a highly challenging animal jumping motion, with and without our proposed adaptive state initialization strategy. As shown in Fig.~\ref{fig:learning_curve}, our method enables effective convergence of policy learning, while the compared one struggles due to physically implausible initial states that hinder the training process.

\paragraph{Impact of Multi-view Guidance}
While our view-agnostic 2D tracking policy leverages diverse training viewpoints to acquire a robust 3D understanding, the inherent depth ambiguity of single-view inputs can still persist during inference for complex motions, causing irregular movements. To investigate this, we conducted an experiment using two orthogonal 2D reference views synthesized by our proposed 2D motion generator. We compared the policy's performance when tracking only one of these views versus tracking both simultaneously. As illustrated in Fig.~\ref{fig:ambiguity_dance}, the policy tracking a single generated view suffers from severe depth ambiguity, resulting in artifacts like foot sliding. In contrast, the policy integrating both views successfully resolves this ambiguity, maintaining stable poses and plausible motion.

\section{Limitation and Future Work}
In our current approach, the 2D tracking policy relies exclusively on reprojection-error-based rewards. This geometric objective may be insufficient for learning precise contact dynamics required for fine-grained object interactions, such as dexterous manipulation or tool use. Future work could address this by introducing additional semantic rewards to explicitly ground these contact constraints.
Moreover, our framework currently assumes that camera parameters are estimated with reasonable accuracy. In practice, however, inaccurate camera estimation can significantly degrade the performance of the 2D tracking policy. To mitigate this, future iterations could incorporate camera parameters directly into the learning procedure, allowing for the joint optimization of motion control and camera estimation.
\end{document}